\documentclass[%
 reprint,
 amsmath,amssymb,
 aps,
]{revtex4-1}
\usepackage{filecontents}

\usepackage{color}
\usepackage{enumerate}

\usepackage{hyperref}
\usepackage{graphicx}
\usepackage{dcolumn}
\usepackage{bm}

\begin{document}


\title{Generative Models for Network Neuroscience: Prospects and Promise}

\author{Richard F. Betzel$^1$}
\author{Danielle S. Bassett$^{1,2}$}
 \email{dsb @ seas.upenn.edu}
\affiliation{
 $^1$Department of Bioengineering, University of Pennsylvania, Philadelphia, PA, 19104}
\affiliation{
$^2$Department of Electrical and Systems Engineering, University of Pennsylvania, Philadelphia, PA, 19104
}

\date{\today}
\begin{abstract}
Network neuroscience is the emerging discipline concerned with investigating the complex patterns of interconnections found in neural systems, and to identify principles with which to understand them. Within this discipline, one particularly powerful approach is network generative modeling, in which wiring rules are algorithmically implemented to produce synthetic network architectures with the same properties as observed in empirical network data. Successful models can highlight the principles by which a network is organized and potentially uncover the mechanisms by which it grows and develops. Here we review the prospects and promise of generative models for network neuroscience. We begin with a primer on network generative models, with a discussion of compressibility and predictability, utility in intuiting mechanisms, and a short history on their use in network science broadly. We then discuss generative models in practice and application, paying particular attention to the critical need for cross-validation. Next, we review generative models of biological neural networks, both at the cellular and large-scale level, and across a variety of species including \emph{C. elegans}, \emph{Drosophila}, mouse, rat, cat, macaque, and human. We offer a careful treatment of a few relevant distinctions, including differences between generative models and null models, sufficiency and redundancy, inferring and claiming mechanism, and functional and structural connectivity. We close with a discussion of future directions, outlining exciting frontiers both in empirical data collection efforts as well as in method and theory development that, together, further the utility of the generative network modeling approach for network neuroscience. 
\end{abstract}

\maketitle
\section*{Introduction}

Many complex systems are composed of elements that interact dyadically with one another and can therefore be represented as graphs (networks) composed of nodes interconnected by edges. The network framework can be applied to systems across a range of disciplines, from sociology and psychology to molecular biology and genomics, making it possible to leverage a common mathematical language and set of analytic tools to investigate the topological organization of systems that, outwardly, might appear dissimilar to one another \cite{newman2010networks}.

In neuroscience, network-based analyses have become common. This is due in part to initiatives for sharing large, multi-modal neuroimaging datasets \cite{nooner2012nki, van2013wu}, the availability of easy-to-use software packages for computing graph-theoretic metrics \cite{rubinov2010complex,sizemore2017dynamic}, and because networks are natural vehicles for representing and analyzing complex spatio-temporal interactions among neural elements (i.e. neurons, populations, and brain areas) \cite{bassett2017network}.

Though the scope of topics studied in network neuroscience is broad, the typical study involves characterizing the structure of a network with a series of summary statistics. Each statistic describes a particular feature of the network, ranging from simple to complex and operating over all topological scales. For example, \emph{degree} is a local (node-level) property that simply counts the number of connections incident upon a node. On the other hand, \emph{characteristic path length} is a global (whole-network) measure of the average length of all pairwise shortest paths. In general, summary statistics offer succinct descriptions of a network's organizational features, especially those that are not immediately apparent given a network's list of nodes and edges.

The application of summary statistics to better understand the structure and function of biological neural networks has been fruitful. Over a decade or so, evidence from networks across different organisms and spatial scales \cite{van2016comparative} has converged onto a small set of properties and summary statistics that, collectively, describe the organization of most biological neural networks. These include indices of small-worldness \cite{bassett2006small}, heavy-tailed degree and edge-weight distributions \cite{hagmann2008mapping, markov2012weighted}, a diverse meso-scale structure that includes segregated modules but also core-periphery structure \cite{sporns2016modular, markov2013cortical}, hubs and rich clubs \cite{zamora2010cortical, van2011rich}, and economic spatial layouts favoring the formation of short-range (low-cost) connections \cite{kaiser2006nonoptimal, horvat2016spatial}. Further, such core organizational principles also include functional constraints, like the need to balance properties that support either the segregated or integrated brain function \cite{sporns2013network}, but also emphasize the tradeoff between the cost of such properties and their functionality \cite{bullmore2012economy}. These properties, collectively, create a caricature of neural system organization and function.

While illuminating, the process of describing networks in terms of their topological properties amounts to an exercise in ``fact collecting." Though summary statistics might be useful for comparing individuals \cite{pena2017spatiotemporal} and as biomarkers of disease \cite{fornito2015connectomics}, they offer limited insight into the mechanisms by which a network functions, grows, and evolves. Arguably, one of the overarching goals of neuroscience (and biology, in general) is to manipulate or perturb networks in targeted and deliberate ways that result in repeatable and predictable outcomes \cite{schiff2012neural}. For network neuroscience to take steps in addressing this goal, it must shift its current emphasis beyond network taxonomy -- i.e. studying subtle individual- or population-level differences in summary statistics -- towards a science of mechanisms and process \cite{vertes2015annual, kaiser2017mechanisms}.

While there exists many methodological approaches for seeking mechanisms in networks and a range of spatial, topological, and temporal scales at which those methods can be deployed \cite{betzel2016multi}, the focus of this article is on network generative modeling. Network generative modeling is a flexible framework for generating synthetic networks from a set of parameterized wiring rules. Generative models figure prominently in the network science canon \cite{erdos1960evolution, wasserman1987stochastic, watts1998collective, barabasi1999emergence}, and have recently been deployed in domain-specific scenarios to study the evolution of protein interaction networks \cite{sole2002model, vazquez2003modeling, middendorf2005inferring}, the world-wide web \cite{kumar2000stochastic}, and social systems \cite{kumar2010structure}. Importantly, and provided that the wiring rule is sufficiently informed and biologically-grounded, generative models can be used to test and identify potential mechanisms that underlie the growth and evolution of biological neural networks. With mechanisms in hand, it becomes possible to distinguish the topological features that drive a network's growth from those that emerge as mere byproducts \cite{rubinov2016constraints}, and to pursue deliberate and targeted interventions \cite{bassett2017networkeng,murphy2017network}.

In the following sections, we present a primer on network generative models, highlighting past use, their interpretation, and open methodological considerations. We review current applications of generative models to neural systems, emphasizing several outstanding questions and implementation details. Finally, we plot a course for future studies.

\section*{Generative Models: A primer}

\begin{figure*}
	\centering
	\includegraphics[width=2\columnwidth]{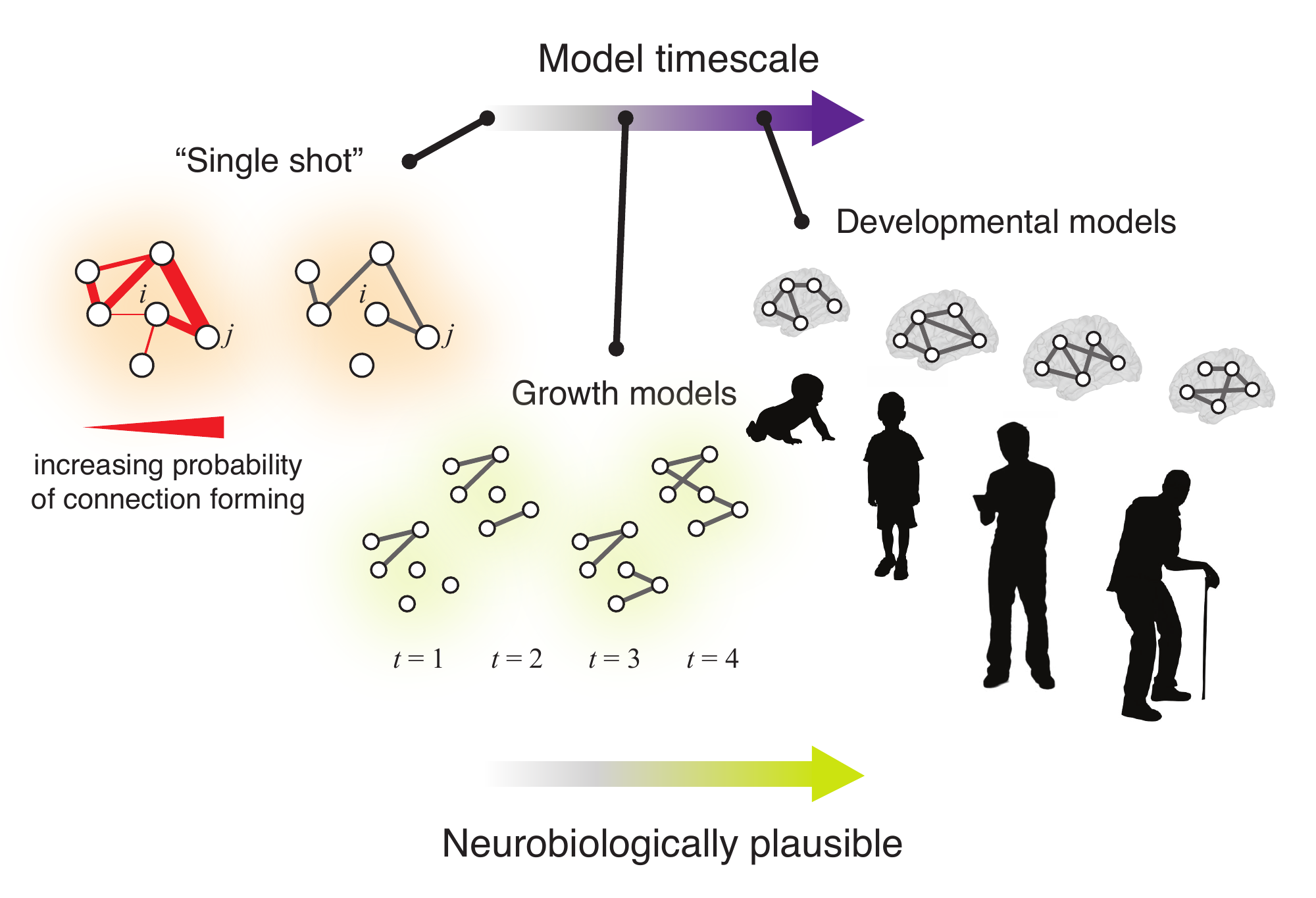}
	\caption{\textbf{Space of generative models.} Generative models can be differentiated from one another along many dimensions, one of which is the timescale over which they operate. A model's timescale is related to its neurobiological plausibility. Models whose timescale is nearer that of developmental time can incorporate more realistic and interpretable features and, in turn, have the chance of uncovering realistic growth mechanisms (e.g. the model of \emph{C. elegans} proposed in \cite{nicosia2013phase}). At the opposite end of the spectrum are ``single shot'' models, e.g. stochastic blockmodels, in which connection probabilities are initialized early on and all connections and weights are generated in a single algorithmic step. Situated between these extremes are growth models that exhibit intrinsic timescales over which connections and/or nodes are added to the network, but where the timescale has no clear biological interpretation.}
	\label{timescalesSchematic}
\end{figure*}

This article deals with the topic of generative models. Broadly, a generative model is a statistical process that outputs a synthetic set of data or observations. Usually, these synthetic data and the generative process are designed to have some properties in common with empirical data and the process believed to have generated those data. Generative models are often parameterized, and those parameters can be chosen so as to minimize the discrepancy between observed and synthetic data. The models, themselves, can be compared against one another using standard model comparison techniques, including goodness-of-fit criteria and cross-validation approaches.

In the context of network science, generative models represent algorithmically-implemented wiring rules that output synthetic networks. While a network's nodes and edges encode all of its structural properties, studying generative models shifts focus away from those structural properties and instead onto wiring rules and the process of network formation. This shift in emphasis confers a number of distinct advantages:
\begin{enumerate}[1.]
	\item Generative models compress our descriptions of networks and highlight regularities in their organization.
	\item They make predictions about out-of-sample and unobserved network data.
	\item Under the best circumstances, generative models can uncover network mechanisms.
\end{enumerate}
We discuss these topics in greater detail throughout the following subsections.

\subsection*{Compressibility of networks}
Generative models compress our descriptions of a network, encoding the network's topology into a set of wiring rules and parameters. Na\"ively, we could describe a network exactly given a list of its nodes and edges, i.e. by consulting the list, we could correctly connect nodes that are supposed to be connected and avoid connecting nodes that should not be connected. However, connections in many networks are not independent of one another and exhibit statistical regularities so that, given the wiring rule that matches those regularities, we could predict the presence/absence of connections ahead of time. In this case, it becomes unnecessary to consult the list of nodes and edges to describe the network. More importantly, we can often interpret the wiring rule itself to uncover the network's organizing principles.

As an example, consider real-world spatial networks, where the probability of observing an edge between two nodes decays as a function of distance \cite{barthelemy2011spatial}. Oftentimes, these kinds of networks can be well-approximated by a simple geometric model whose wiring rule mimics the network's distance-dependent connection formation \cite{dall2002random}. To perfectly describe a spatial network we could generate a long and possibly unwieldy list of its nodes and edges. However, if the geometric model is a good approximation, e.g. synthetic networks generated by the model recapitulate many observed edges, then the model can be used to replace those edges in the list, effectively shortening our description of the network. The geometric model naturally mimics the distance dependencies of the spatial network. For many networks, however, the statistical regularities among links may not be obvious, in which case selecting the appropriate model may not be straightforward. We discuss this issue of model selection later in this section.

\subsection*{Predictability}
Besides compressing our descriptions of a network, generative models also have predictive capacity and can be used as forward models of unobserved and out-of-sample data. Returning to the example of spatial networks, we might hypothesize the relevance of a generative model in which the probability of connection formation is given by a decaying exponential. If we let $A_{ij} \in \{0,1\}$ indicate the presence or absence of an edge between nodes $i$ and $j$, we can write this connection probability as: $P(A_{ij} = 1) \propto \exp(-\beta \cdot D_{ij})$, where $D_{ij}$ is the distance between nodes $i$ and $j$ and $\beta \ge 0$ is a parameter to be fit \cite{kaiser2004spatial}. If we were given a network $G$, we could fit the parameter $\beta$ so that the discrepancy between synthetic networks generated by the model and $G$ is minimized. Having fit the model, we could use it to make predictions about a second network, $G^\prime$, whose connectivity pattern is unknown but whose nodes' spatial locations are given.

As another example, consider the stochastic blockmodel \cite{wasserman1987stochastic,snijders1997estimation}, in which nodes are assigned membership to one of $K$ communities, $z_i \in \{1,\ldots,K\}$, and where the probability of two nodes, $i$ and $j$, being connected to one another depends only on their community assignments: $P(A_{ij} = 1) = \omega_{z_i,z_j}$ ($\omega$ is a $K \times K$ matrix that encodes community-to-community connection probabilities). Fitting this model to a network $G$ entails inferring nodes' communities and connection probabilities. If we encountered a second network, $G^\prime$, with an unknown connectivity pattern but whose nodes correspond to those in $G$, e.g. the same set of neurons or brain regions, then we could use the model to predict the configuration of nodes and edges in that network.

\subsection*{Mechanisms}

Finally, provided that it incorporates sufficient system-specific details (in our case, neurobiological information), a generative model can be used to gain insight into the mechanisms that guide the formation and growth of a system. This last point is critical. A generative model, under ideal circumstances, is a recipe for building a network. Having such a recipe opens new avenues for interrogating a network. It allows us to identify structural features of a network that emerge as a direct result of the wiring rule, \emph{versus} those that emerge spontaneously as a consequence of constraints imposed by a given wiring rule \cite{rubinov2016constraints}. For example, a geometric model will generate networks with high levels of clustering even though the wiring rule never explicitly optimizes for this property. Importantly, a recipe for building a network also gives us the ability to explore alternative ingredients. What happens if we change a parameter slightly? Does the model generate networks of vastly different character? Can we control the trajectory of a network's growth and guide it into a desired target configuration \cite{gu2015controllability}? The ability to selectively drive the growth of a network is a tantalizing prospect, and one with profound implications in the treatment of clinicial and psychiatric disorders.

\subsection*{Canonical generative models for networks}

Before engaging neuroscience-specific questions, it is useful to discuss examples of generative models as they have been applied in network science and other fields. In the remainder of this section we review some canonical generative models, emphasizing the properties that they share with one another as well as those that make them distinct.

Generative models have a long history in network science and mathematics. One of the earliest examples is the so-called Erd\H{o}s-R\'{e}nyi (ER) model \cite{erdos1960evolution}, in which connections are formed independently between pairs of $N$ nodes with probability $P$ (another version exists where, instead of $P$, a fixed number of edges, $M$, are added uniformly at random). While the ER model has interesting combinatoric and mathematical properties, e.g. binomially-distributed node degree \cite{bollobas1998random}, it is a poor approximation of most real-world networks. That is, the random and independent process by which connections are formed in the ER model results in networks with no real structure (poor compressibility) and does not resemble any of the mechanisms by which real-world networks grow. Accordingly, if we want to model networks in the real world, we need a set of models that generate networks with realistic properties.

Initial explorations into generative models for real-world data resulted in two models that, collectively, helped spark broad interest in complex networks. The first, introduced by Duncan Watts and Steven Strogatz, sought the origin of empirically observed ``small world'' topologies, in which a network simultaneously exhibits greater-than-expected clustering and shorter-than-expected path length \cite{watts1998collective}. Broadly speaking, the model supposed that small-world networks are an interpolation between two extreme configurations: a ring lattice network (nodes arranged on the circumference of a circle and linked to their $k$ clockwise and counter-clockwise neighbors) and an ER network. To move from one extreme to the other, the authors introduced a tuning parameter, $p$, which governed the probability that an edge in the lattice network would be rewired randomly. When $p$ is small, the model generates networks that have mostly lattice-like properties, but when $p$ is large, the model generates networks whose properties are indistinguishable from those produced by the ER model. Between those extremes, however, is a ``sweet spot'' -- a region of parameter space yielding networks with properties of both extremes, namely high clustering and short path length. This model is referred to as the Watts-Strogatz (WS) model.

At around the same time, a second group sought an explanation for why many real-world networks exhibited heavy-tailed degree distributions. The proposed model, by R\'{e}ka Albert and Albert-L\'{a}szl\'{o} Barab\'{a}si, was based on a growth rule \cite{barabasi1999emergence}. Starting with a small set of fully-connected nodes, the model adds new nodes to the network by forming connections preferentially to already-existing nodes with higher degrees. This growth mechanism is a sort of ``rich get richer'' process; nodes that have existed for a long time accumulate many connections, which further increases their likelihood of being connected to newly-added nodes. The result of this process is a network with an approximately power-law degree distribution, mimicking those frequently observed in real networks \cite{clauset2009power}. This model is identical to that defined by Price in 1976 with a single value change to one parameter \cite{price1976general}, and is generally referred to as the Barab\'{a}si-Albert (BA) or preferential attachment (PA) model.

\subsection*{Generative models in practice and application}

The WS and BA models generate synthetic networks with properties \emph{qualitatively} similar to those observed in real-world networks (small-worldness and heavy-tailed degree distribution). If we wanted to make the similarity of empirical and synthetic networks quantitative and more precise, how would we do so? Supposing that a model yields networks that repeatably and exactly recapitulate \emph{all} properties of an empirical network, can we equate the model with mechanism? Both of these questions are difficult to answer, and represent some of the technical challenges associated with generative modeling.

\subsubsection*{Choosing an objective function}

We will first address the issue of how to perform quantitative comparisons between synthetic and empirical networks. Fortunately, there exists a plurality of approaches for quantitatively comparing networks. The challenge is selecting the approach that is best suited for a given research question.

Typically, we wish to answer the question of whether an empirically-observed network could have been produced by some generative model. One strategy for addressing this question involves defining a likelihood function over the space of all possible networks, and evaluating that function for the observed network. Stochastic blockmodels are a good example of this strategy in action \cite{holland1983stochastic, wasserman1987stochastic, snijders1997estimation}. The probability of a connection forming between nodes $i$ and $j$, $P_{ij}$, depends on their community assignments, $z_i$ and $z_j$: $P(A_{ij} = 1) = \omega_{z_i,z_j}$. The probability that $i$ and $j$ are disconnected, is therefore $1 - P_{ij}$ and the likelihood that the observed network was generated by this model is given by:

\begin{equation}
\mathcal{L} = \prod_{i,j > i} P_{ij}^{A_{ij}} (1 - P_{ij})^{1 - A_{ij}}.
\end{equation}

\noindent Blockmodels are convenient in that this likelihood function can be written in closed form. This approach can be generalized for other models -- even when the precise likelihood function is unknown -- by generating a sample of networks from a given step of parameters and estimating, from those samples, the probability of any connection existing.

This approach is similar to others in the literature \cite{costa2007predicting}, in that it links the model's fitness with its ability to correctly account for the empirical network's exact configuration of nodes and edges. While this seems like a good approach, it is not difficult to envision scenarios where even near-perfect prediction of an empirical network's connections nonetheless fails to account for some of its critical topological properties. For example, consider the canonical small-world network -- a ring lattice plus a few random (shortcut) connections that reduce the network's characteristic path length. The ring lattice and small-world network have nearly-perfect edge overlap. If we were to regard edge overlap as the definitive measure of fitness, we might be inclined to treat the lattice network as a good approximation of the small-world network. In other words, from a strictly structural point of view, these two networks are almost perfect matches; from a functional perspective, however, the two networks are highly dissimilar; because of its longer characteristic path length the ring lattice will lack efficient (short) routes possible used for communication or transportation.

Comparing synthetic and empirical networks on the basis of their edge configuration is useful, but has some shortcomings that motivate the exploration of alternative approaches.  Another approach, and one that has been used in several recent studies \cite{vertes2012simple, betzel2016generative}, eschews the edge-wise comparison of two networks, instead simultaneously comparing them along several topological dimensions (e.g., their efficiency, clustering, modularity, etc.), and calculating a statistic of average dissimilarity. For example, in V\'{e}rtes et al. (2012), the authors compared synthetic and empirical networks with the energy function, $E = \frac{1}{V_C \cdot V_E \cdot V_M \cdot V_K}$, where $V_C$, $V_E$, $V_M$, and $V_K$ are $p$-values associated with statistical tests comparing clustering, efficiency, modularity, and degree distribution of synthetic and empirical networks \cite{vertes2012simple}. Similarly, Betzel et al. (2016) defined the energy function $E = \max(\text{KS}_K,\text{KS}_C,\text{KS}_B,\text{KS}_E)$, where each term is a Kolmogorov-Smirnov statistic comparing the degree ($K$), clustering ($C$), betweenness centrality ($B$), and edge length ($E$) distribution \cite{betzel2016generative}. Intuitively, in both cases smaller energies imply greater fitness.

This approach is flexible and can be adapted to include virtually any set of metrics. It is important to note, however, that many network measures are correlated with one another, so the choice of which to include should take this into account. Also, there might be synthetic networks that match an empirical network in terms of network statistics but not its precise set of connections. Irrespective of how the objective function is defined, having one makes it possible to perform different kinds of comparisons. For a given model, we can perform model fitting by selecting the parameter values that optimize the objective function. We can also leverage an objective function to compare different generative models to one another. For example, we may wish to discount a model that is incapable of generating networks that resemble our real-world network of interest.

\subsubsection*{Cross-validation}

Suppose that we fit a generative model by optimizing some objective function so that the model generates synthetic networks that share some set of properties with an empirical network. As in any model-fitting exercise, we can continue adding layers of complexity and free parameters to the model so that it matches our real-world network to some arbitrary degree of precision. It is often the case, however, that we are less interested in predicting the organization of a single network, but of a class of networks. For example, we may wish to identify wiring rules that can recapitulate the organization of structural brain networks, on average, rather than the network of any one individual. Even if our aim was to predict subject-specific networks, it might be unsurprising (in a statistical sense) that our models reproduce many of the features of those networks; after all, the model's parameters were selected only after an optimization procedure.

In both cases (fitting models to empirical network data based on edge- or property-matching), it is essential that we perform a cross-validation procedure. This procedure might entail taking the best-fitting parameters from one model and using them to generate estimates of a second network not involved in the model-fitting process. We can compare the goodness of fit to that of a random (ER) model, to ensure that our model performs above chance. This type of cross-validation ensures that a generative model is identifying general wiring rules and not overfitting. A second type of cross-validation involves testing whether synthetic networks have properties in common with real-world networks that they were not explicitly optimized to possess. In other words, does a generative model give us certain properties ``for free?'' This type of cross-validation ensures that our objective function is sufficiently general and not overfitting and emphasizing a specific subset of network properties.

\subsection*{The space of generative models}

What distinguishes one generative model from another? Is it possible to delineate classes of generative models based on their functions or characteristics? Arguably, one of the distinguishing features of any generative model is the timescale over which it operates (Fig.~\ref{timescalesSchematic}). On one extreme are models with no timescale at all, like stochastic blockmodels \cite{snijders1997estimation,wasserman1987stochastic}. These kinds of models are ``single-shot'' generators of networks, and can therefore be quite poor representations of real-world networks that grow and evolve over time. On the other extreme are models whose internal timescale matches that of the real system. Nodes and edges are added or rewired on a realistic timescale to match known properties of the system. The growth model of \emph{C. elegans} presented by Nicosia et al. (2013) is a good example \cite{nicosia2013phase}. In this model, nodes and edges are added according to their empirically measured birth times (time of cell division); a feature that contributed to the success of that model in predicting different properties of the \emph{C. elegans} connectome.

Between these two extremes -- where models operate with either no intrinsic or biologically plausible timescale -- is where most generative models are situated. In this middle ground, edges and nodes are added to or rewired in an existing network, but the timescale over which these processes occurs is arbitrary. A good example is the BA model, in which new nodes are linked to an existing network over a series of steps. These steps are ordered, so the addition of one node precedes or follows that of another. However, time is measured in arbitrary units (steps) and direct comparison to biological timescales, e.g. human development, might be inappropriate. Ordering generative models based on their internal timescales is similar to ordering them according to their plausbility and mechanistic understanding. Blockmodels and models with arbitrary timescales can do a good job compressing our description of a network and may identify general organizational principles \cite{rubinov2016constraints}. However, if our aim is to develop realistic mechanistic models of network growth and development, it is essential that we include the necessary components that ground the model in reality.

\section*{Generative models of biological neural networks}

Now that we have an intuition for what a generative model is, and what the goals are for building a generative model, we turn to a brief review of existing generative models for biological networks observed in neural systems. We note that this review is not comprehensive, but instead focuses on areas in which significant work has been accomplished, or areas that motivate important current and future frontiers. We also refer readers elsewhere for additional details on the mechanisms of connectome development \cite{kaiser2017mechanisms}, biophysical models of neural dynamics \cite{breakspear2017dynamic}, and modeling mesoscale structure in dynamic networks \cite{khambhati2017modeling} and multiscale networks \cite{betzel2016multi}.

Finally, we note that this review focuses mostly on generative models of structural and not functional networks (the distinction is in how edges are defined; in structural networks they represent physical connections, e.g. synapses, projections, fiber tracts, whereas in functional networks they represent statistical associations among neural elements' activity, e.g. correlation, coherence, etc.). Because of differences in how structural and functional networks are generated and evolve, certain classes of models that are appropriate for one type may be wholly inappropriate for the other. For example, functional networks are not generated through an edge addition process -- they emerge from constrained dynamical processes. We discuss the implications of these differences in more detail later in this section.

\subsection*{The requisite ingredients}

An open and important question that scientists face when embarking on a study to develop a generative model is: ``What features are required to build good network models?'' Perhaps the simplest feature one requires is a target network topology, the organization of the network that one is trying to recapitulate and ultimately explain. Yet, a single network topology can be built in many different ways, with strikingly different underlying mechanisms \cite{stumpf2012mathematics}. Thus one might also wish to have a deep understanding of (i) the contraints on anatomy, from physical distance \cite{karbowski2001optimal} to energy consumption \cite{laughlin1998metabolic}, (ii) the rules of neurobiological growth, from chemical gradients \cite{borisyuk2014developmental} to genetic specification \cite{kennedy1993importance}, and (iii) the pressures of normal or abnormal development, and their relevance for functionality. Moreover, each of these constraints, rules, and pressures can change as the system grows, highlighting the importance of developmental timing \cite{kennedy1993importance}. Of course, one might also wish to choose which of these details to include in the model, with model parsimony being one of the key arguments in support of building models with fewer details.

\begin{figure*}
	\centering
	\includegraphics[width=2\columnwidth]{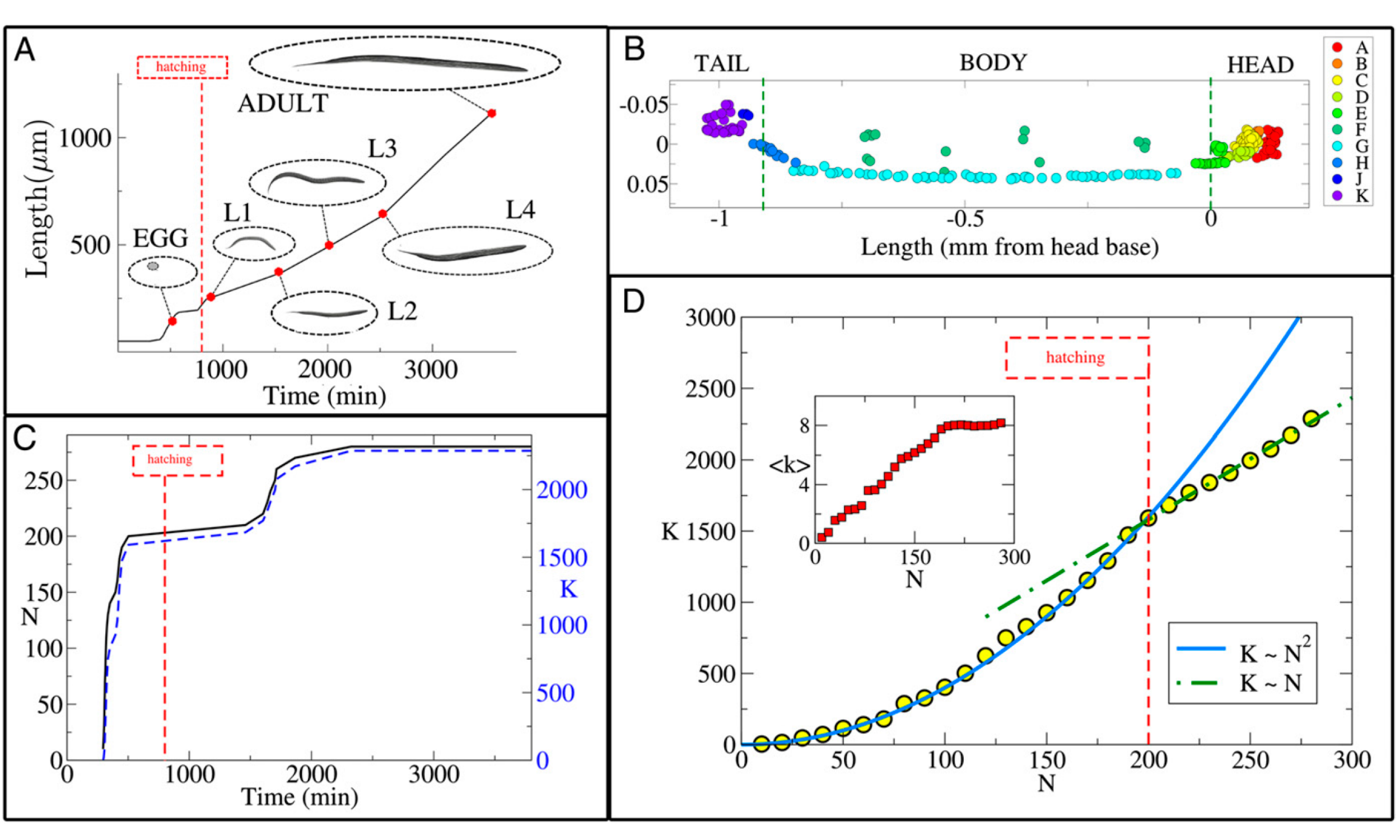}
	\caption{\textbf{Development of the \emph{C. elegans} nervous system.} \emph{(A)} \emph{C. elegans} reaches adulthood approximately 63 hours after fertilization, over which time its body increases appreciably in length. \emph{(B)} In the adult hermaphrodite worm, neurons are distributed unevenly across the body, with more than 60\% being located in the head and about 15\% being located in the tail tip. Here, neurons are color-coded according to their membership to the following ganglia: anterior [A], dorsal [B], lateral [C], ventral [D], retrovesicular [E], ventral cord [G], posterior lateral [F], preanal [H],
dorsorectal [J], and lumbar [K]. \emph{(C)} The total number of neurons ($N$, solid black), and connections ($K$, dashed blue), grows nonlinearly but monotonically with time. \emph{(D)} A phase transition is evident in the number of synapses as a function of the number of neurons (yellow circles): before hatching, $K$ grows as $N^{2}$ (solid blue line), whereas after hatching, $K$ grows linearly with $N$ (dashed green line). (Inset) Plot of the average nodal degree, $K$, \emph{versus} number of nodes, $N$. Reprinted with permission from \cite{nicosia2013phase}.}
	\label{fig:nicosia}
\end{figure*}

\subsection*{Generative models at the cellular level}

Recent efforts to model cellular level network architecture have had the benefit of building on rich empirical observations made over the last several decades. At one of the smallest spatial scales of neuronal connectivity, evidence suggests that the arbors of single neurons can be characterized by both local \cite{cherniak1992local} and global \cite{cherniak1999large} optimization rules to more strongly minimize volume than length, signal propagation speed, or surface area. Within the confines of relative volume cost minimization, there is also evidence for a maximization of the repertoire of possible connectivity patterns between dendrites and surrounding axons: in basal dendritic arbors of pyramidal neurons, arbor size scales with the total dendritic length, the spatial correlation of arbor branches appears to have a single functional form, and small sections of an arbor display self-similarity \cite{wen2009maximization}. 

The morphology of dendritic arbors specifically and other parts of the cell more generally have direct bearing on the degree of connectivity that can take place between neurons \cite{chklovskii2004synaptic}. Like dendritic arbors, synaptic connectivity appears to be organized in a highly non-random manner \cite{song2005highly}, with unexpectedly high density in relation to its volume \cite{chklovskii2004synaptic}. Interestingly, both synaptic connectivity and neuronal morphology appear to experience some similar constraints, including principles of wiring optimization \cite{cherniak2012neural,karbowski2001optimal}. Some suggest that constraints on synaptic wiring may be the more fundamental of the two, explaining the degree of separation between cortical neurons \cite{karbowski2001optimal}, as well as the placement of cell bodies \cite{rivera2014wiring}. Others suggest that it is in fact the combination of wiring economy and volume exclusion that can determine neuronal placement \cite{chklovskii2011wiring}. 

In either case, the highly non-random nature of synaptic connectivity has been the subject of several recent generative modeling efforts. Initial observations that this non-random organization could be parsimoniously described as small-world \cite{bassett2006small,bassett2016small}, have motivated the question of how this particular type of network complexity is combined with pressures for wiring minimization. Nicosia et al. (2013) suggest that the growth rules shaping cellular nervous systems balance an economical tradeoff between wiring cost and the functionality of network topology (Fig.~\ref{fig:nicosia}). Using a dynamic economical model incorporating a continuously negotiated tradeoff between wiring cost and network topology, they recapitulate an empirically observed phase transition in the proportion of nodes to links present over the developmental time period of \emph{C. elegans} \cite{nicosia2013phase}. The authors speculate that such dynamically negotiated tradeoffs may be characteristic of other complex systems, whether biological or manmade. It will be interesting in the future to consider scenarios in which such tradeoffs may be negotiated over shorter time periods, such as in the alteration of the prevalence of autaptic connections posited to play a role in homeostatic network control of bursting \cite{wiles2017autaptic}. 

The incorporation of a dynamic economic tradeoff is an example of the broader importance of incorporating biophysically accurate features in generative models of cellular neural systems. Another example of such a biophysical feature is axon and dendrite geography, which has been shown to predict the specificity of synaptic connections in a functioning spinal cord network of hatchling frog tadpoles \cite{li2007axon}. Some generative models have also sought to determine the role of neuron type in observed network topology and function, for example by building models of sensory neurons, sensory pathway interneurons, central pattern generator (CPG) interneurons, and motoneurons, and then linking them in a network with known inter-type connectivity \cite{sautois2007role}. By adding knowledge about development including chemical gradients and physical barriers \cite{borisyuk2014developmental}, a cell-type specific model of 2000 neurons in the spine of a young Xenopus tadpole can produce swimming behavior in response to sensory stimulation \cite{roberts2014simple}. These and related efforts demonstrate the ability of generative network models built with neuron and synapse resolution, and incorporating biophysical phenomena, to reproduce behaviors observed in whole organisms. Such findings are reminiscent of other biophysical modeling efforts at the large scale of human areal networks \cite{jones2009quantitative,jones2016when}, where the biophysics of regional rhythms and inter-regional synchronization inform our understanding of human cognition \cite{kopell2014beyond}.

Of course, statistically bridging structural connections such as synapses at the cellular scale with behaviors in non-human animals -- and cognition in humans -- at the organism scale begs the question of what processes exist between the two scales.  There do exist generative models of functional network topology from structural network topology, and \emph{visa versa}.  A particularly powerful approach for cellular nervous systems is the pairwise maximum entropy model \cite{chau2017inverse} and recent extensions \cite{stein2015inferring}, which can be used to predict patterns of pairwise correlations from structure, or to infer structure from pairwise correlations. This latter inference neglects unmeasured higher-order (non-pairwise) interactions, operationalized via the maximum entropy distribution, which assumes maximal independence among variables (in this case: cells) \cite{tang2008maximum,shlens2006structure}. The technique was initially applied to neural spiking data to demonstrate that, in the case of the energy function being the Ising model, pairwise interactions give an excellent approximation of the full correlation network \cite{scheidman2006weak}. The surprisingly good fit of this model to the data has important implications for how we think about neural population codes in response to stimuli \cite{schneidman2016towards}, which can be represented by joint activity patterns of spiking and silence \cite{granot2013stimulus,cofre2014exact}. Moreover, the maximum entropy model also provides a surprisingly accurate fit to large-scale imaging data in the form of fMRI BOLD collected in humans at rest \cite{watanabe2013pairwise}, as well as during a task \cite{watanabe2014energy}. Interestingly, the simple assumptions of this minimal generative model also appear to provide excellent fits to the dynamics of mesoscale network communities in functional data \cite{ashourvan2017energy} and insights into the energy landscape that the system traverses \cite{ezaki2017energy}.

While the pairwise maximum entropy model has proven useful in inferring structural network organization from functional network organization, and \emph{visa versa}, it is certainly true that non-pairwise interactions may nevertheless play a non-trivial role in neural population function. Intuitively, beyond-pairwise interactions can occur via common input \cite{barreiro2014when}, producing multi-way synchrony \cite{kelly2012framework} with varying prevalence across different length scales in the system \cite{ifije2010sparse}. Generative models of such high-order relations include beyond-pairwise maximum entropy models \cite{ganmor2011sparse} and dichotomous Gaussian models \cite{leen2015simple}. Another way in which beyond-pairwise functional interactions can occur is if neurons themselves do not only display pairwise connections, but also higher-order connections. This possibility higlights a complementary challenge in describing the presence of such higher-order relations in structural networks from a topological point of view, with the goal of building generative models that account for them. 

A useful language with which to meet this challenge is the language of algebraic topology and specifically simplicial complexes \cite{giusti2016twos} whose fundamental units are simplices: a 0-simplex is a node, a 1-simplex is a dyad, a 2-simplex is a face, a 3-simplex is a tetrahedron, a 4-simplex is a 5-cell, etc. A collection of simplices -- called a simplicial complex -- can include many interesting features including cliques (i.e., fully connected subgraphs) and cavities (collections of $n$-simplices arranged so that they have an empty geometric boundary). In patterns of correlations among the activity of pyramidal neurons in rat hippocampus, the topology of cliques and cavities demonstrates geometric organization consistent with a generative model of simplicial complexes related to random geometric graphs \cite{giusti2015clique}. This higher-order structure has also enabled the identification of unexpectedly long structural loops linking regions of early and late evolutionary origin, underscoring their unique role in controlling brain function \cite{sizemore2016cliques}. Indeed, the topology of cliques and cavities has specific implications for local processing (cliques) \emph{versus} processing in which information may flow in either diverging or converging patterns (cavities) \cite{sizemore2016cliques}, and can support efficient coding by enabling inference of neural codes even in highly undersampled set of patterns \cite{curto2013neural}. While generative models of simplicial complexes based on random geometric graphs have shown some utility in explaining these structures, further work is needed to understand the extent of their applicability, and to consider models for growing simplicial complexes \cite{courtney2017weighted}. 

\begin{figure*}
	\centering
	\includegraphics[width=2\columnwidth]{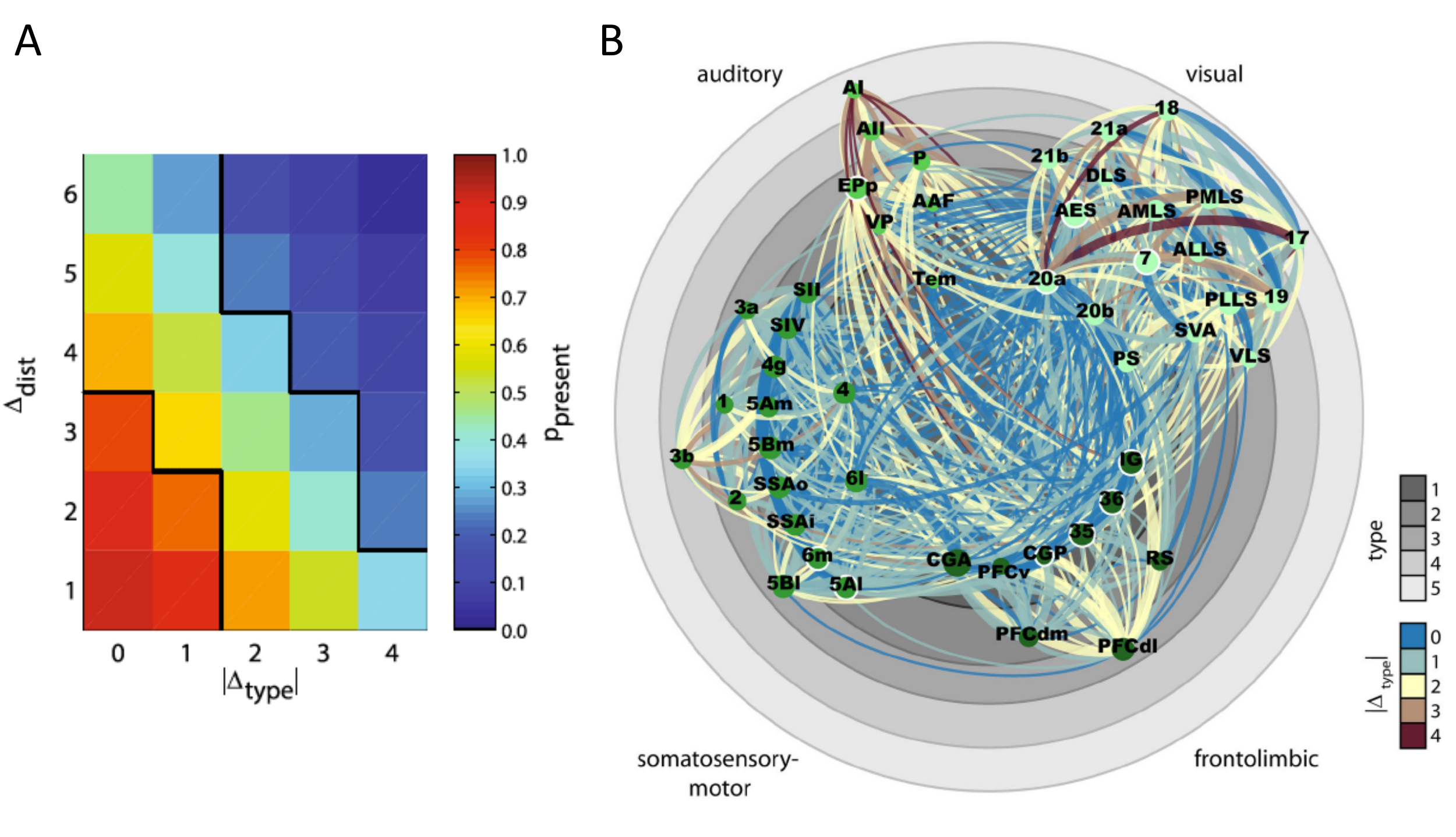}
	\caption{\textbf{A predictive model of the cat cortical connectome based on cytoarchitecture and distance.} \emph{(A)} Results of linear discriminant analysis showing posterior probabilities ($p$) for presence of projections across the 2-dimensional predictive variable space defined by the border distance between regions ($\Delta_{\mathrm{dist}}$) and the absolute type difference between regions ($| \Delta_{\mathrm{type}} |$). Black borders enclose ranges of $p$ present $> 0.75$ and $p$ present $<0.25$. \emph{(B)} Visualization of the corticocortical connections collated in Scannell et al. (1995) \cite{scannell1995analysis}. All present projections between cortical areas for which a structural type was defined (49 of 65 areas) are displayed. Circles correspond to structural types, cortical areas are placed accordingly. Structural type increases from center to periphery. Projections are color-coded according to the absolute structural type difference of the connected areas. Ordinal projection strength (sparse, intermediate, or dense) is coded by increasing projection width. Nodes are grouped and color-coded according to anatomical modules as indicated. Node sizes indicate the areas’ (unweighted) degree. Hub-module areas, as classified by Zamora-L\'{o}pez et al. (2010) \cite{zamora2010cortical}, are marked by a white outline. Reprinted with permission from \cite{beul2015predictive}.}
	\label{fig:beul}
\end{figure*}

\subsection*{Increasing in scale: generative models of large-scale connectomes}

In the previous section, we reviewed some of the literature supporting the notion that cellular network organization in neural systems is characterized by pressures of wiring economy and topological complexity. Such pressures are similarly thought to play a role in the organization of networks at the meso- and large-scale in both human and non-human mammalian brains \cite{bullmore2012economy}. Computational studies suggest that trade-offs between wiring economy and topological complexity \cite{tononi1994measure} support the formation of network modules, offering relative segregation of function, and network hubs, offering relative integration of function \cite{chen2013tradeoff}. The role of topological complexity and the presence of unusually high wiring costs in some parts of cortex suggests that simple notions of spatial embedding are not sufficient to explain the observed organization of the connectome. This limitation has motivated models deriving a latent (rather than physical) space from which to predict missing links \cite{hinne2017missing}, or incorporating information about cytoarchitecture \cite{beul2015predictive} such that cytoarchitectonically similar cortical areas in the two hemispheres have an unexpectedly high probability of connecting with one another \cite{goulas2017principles}. 

A particularly salient example of a generative model of areal connectivity in a mammalian brain that incorporates many of these considerations is the recent predictive model of Beul et al. (2015) (Fig.~\ref{fig:beul}). In this paper, the authors study meso-scale structural connectivity between 49 areas of the cat cerebral cortex as estimated by tract tracing techniques \cite{beul2015predictive}. They test the predictive utility of 3 separate wiring rules: (i) a \emph{structural} rule in which the laminar patterns of origins and terminations of inter-areal projections vary according to the relative cytoarchitectonic differentiation of the projection sources and targets, (ii) a \emph{distance} rule in which connections are more frequent, and more dense, among neighboring regions and sparser or absent between remote regions, and (iii) a \emph{hierarchical} rule in which differences in the functional hierarchical levels of source and target areas are inversely related to the degree of connectivity between them. While the latter rule did not accurately fit the data, the first two rules (structure and distance) explained significant variance in the observed connectivity patterns, with a linear combination of the two predicting the existence of connections with more than 85\% accuracy.
 
Work in non-human primates generally and the macaque cortex specifically recapitulates many of the same motifs from work in lesser mammals. Early work suggested that cortical components are optimally placed so as to minimize the costs of their interconnections \cite{cherniak1994component}, facilitating a global optimal cerebral cortex layout \cite{cherniak2004global}. Later work suggested that component placement did not maximally minimize wiring, but also tended to favor short processing paths, due to long-distance projections \cite{kaiser2006nonoptimal}. Indeed, separate from where components are placed, it has been noted that there appear to be successfully arbitrated optimization problems in the organization of inter-areal connectivity, for example favoring near-minimization of distance \cite{horvat2016spatial,ercsey2013predictive} and increased support for connectivity between areas with similar topological properties \cite{costa2007predicting}. In an extension of the model described above for the cat, Beul and colleagues similarly demonstrate the striking utility of the structural rule of architectonic similarity, where similarity in the laminar pattern of projection origins, and the absolute number of cortical connections of an area, demonstrated the strongest and most consistent influence on connection features \cite{beul2017predictive}. In this case, the distance rule was surprisingly not predictive. Future extensions of this model may include explicit nonlinear growth rules, which have previously been linked to the emergence of network hubs \cite{bauer2017nonlinear}. 

Finally, efforts in the human support the notions of wiring economy \cite{raj2011wiring,samu2014influence} and topological complexity \cite{betzel2016generative}, and further add new considerations such as the geometric segregation of the brain into gray and white matter, enabling the relative minimization of conduction delays \cite{wen2005segregation}. While one-shot models have been the most commonly exercised generative models for human structural networks, relatively new evaluation criteria for them include an assessment of their controllability profiles \cite{wuyan2017benchmarking} and homological features \cite{sizemore2017classification}. Moreover, there has been a recent and growing interest in developing network growth models that incorporate biologically motivated rules for the probability of connections \cite{klimm2014resolving,jarman2014spatially}. For example, spatially constrained adaptive rewiring creates small-world network architectures with spatially localized modules \cite{jarman2014spatially}, while wiring rules based on topological affinities recapitulate known scaling laws of physical network topology \cite{klimm2014resolving}. It would be interesting in future work to determine how these rules could be adapted to explain the patterns of conserved and variable architecture of white matter networks across individual humans \cite{bassett2011conserved}.

\begin{figure*}
	\centering
	\includegraphics[width=1.75\columnwidth]{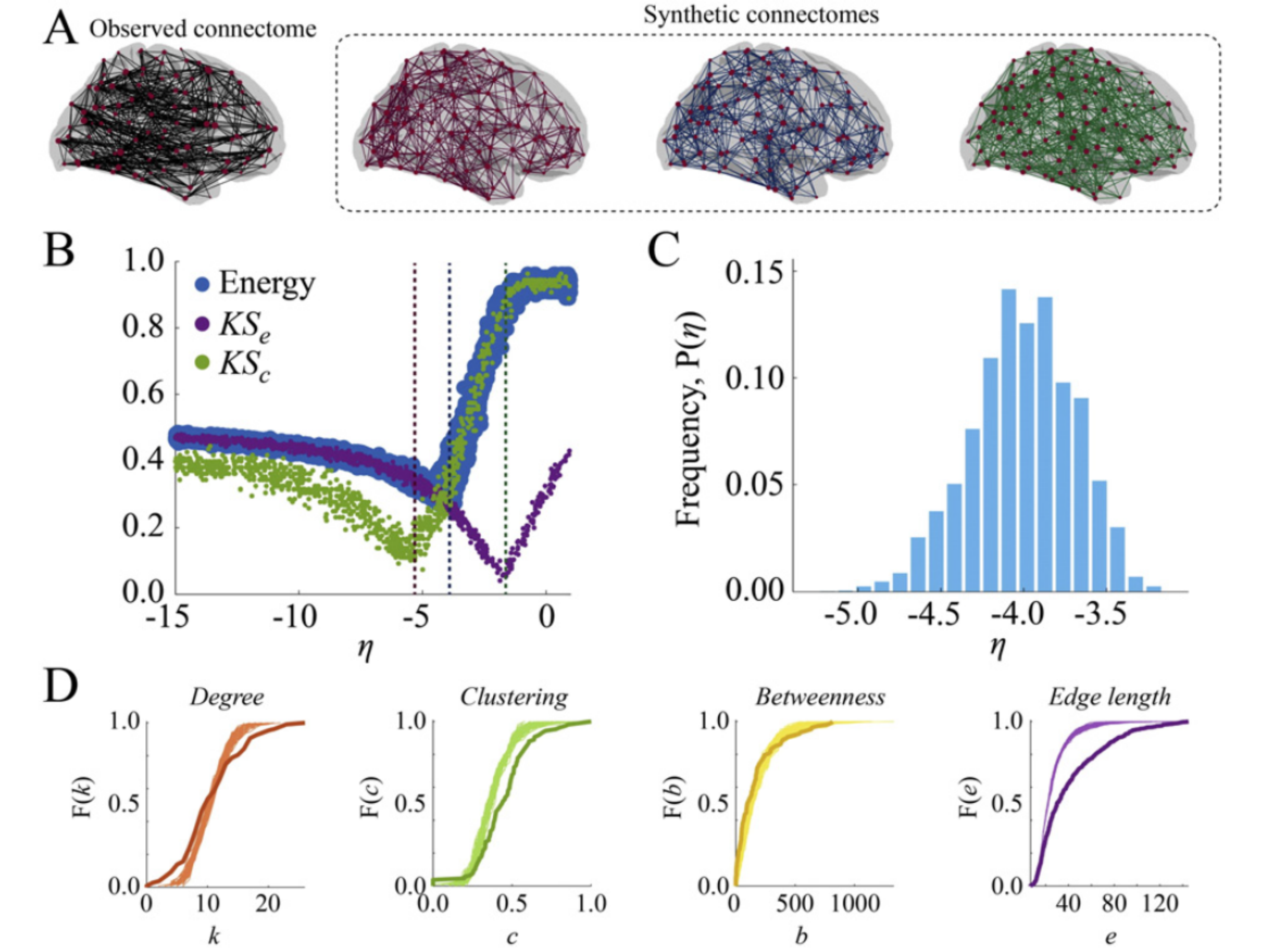}
	\caption{\textbf{Summary of a geometric model for human white matter networks estimated from diffusion imaging.} \emph{(A)} Observed (black) and synthetic (colors) networks generated at different points in a pre-defined parameter space of interest. \emph{(B)} Energy landscape where the energy function is defined as $E = \max(\text{KS}_K,\text{KS}_C,\text{KS}_B,\text{KS}_E)$, where each term is a Kolmogorov-Smirnov statistic comparing the degree ($K$), clustering ($C$), betweenness centrality ($B$), and edge length ($E$) distribution between the observed and synthetic networks \cite{betzel2016generative}. Specifically, here we show the behavior of $KS_{e}$, $KS_{c}$, and energy as a function of $\eta$, a parameter that controls the characteristic connection length such that when $\eta < 0$, short-range connections are favored, while when $\eta > 0$ long-range connections are favored. The dashed vertical lines indicate the parameter values at which the example synthetic networks were generated. \emph{(C)} Distribution of $\eta$ parameter of top 1\% lowest-energy synthetic networks aggregated across all  human participants in the study. \emph{(D)} Cumulative distributions of degree (orange), clustering coefficient (green), betweenness centrality (yellow), and edge length (purple) for observed connectome (darker line) and best-fitting synthetic networks (lighter lines) for a representative participant. Reprinted with permission from \cite{betzel2016generative}.}
	\label{fig:betzel}
\end{figure*}

The recent paper by Betzel et al. (2016) represents one of the first attempts at subject-level generative modeling \cite{betzel2016generative}. In this study, the authors fit thirteen generative models to white-matter networks acquired from three independent datasets, totaling 380 subjects (Fig.~\ref{fig:betzel}). The model generated synthetic networks using an edge-addition algorithm, in which connections were added probabilistically and one at a time according to a set of parameterized wiring rules. Each of the thirteen models was fit in two stages: first by matching distributional statistics of the white-matter networks and later cross-validated on a separate set of network measures. The best-fitting models across all three datasets featured wiring rules based on wiring cost reduction and homophilic attraction principles, the severity of each controlled by a separate parameter. Because the models were fit to individual subjects, it was possible to explore individual variability in model fit. When applied to lifespan data from the Nathan Kline Institute, the authors found that the parameter governing the severity of the wiring cost reduction weakened systematically with age, as did the model goodness of fit. These findings suggest that generative models are sensitive to changes in network organization with development and aging, and may be useful tools in studying variation across individuals \cite{zuo2016human}.

\begin{figure*}
	\centering
\includegraphics[width=1.75\columnwidth]{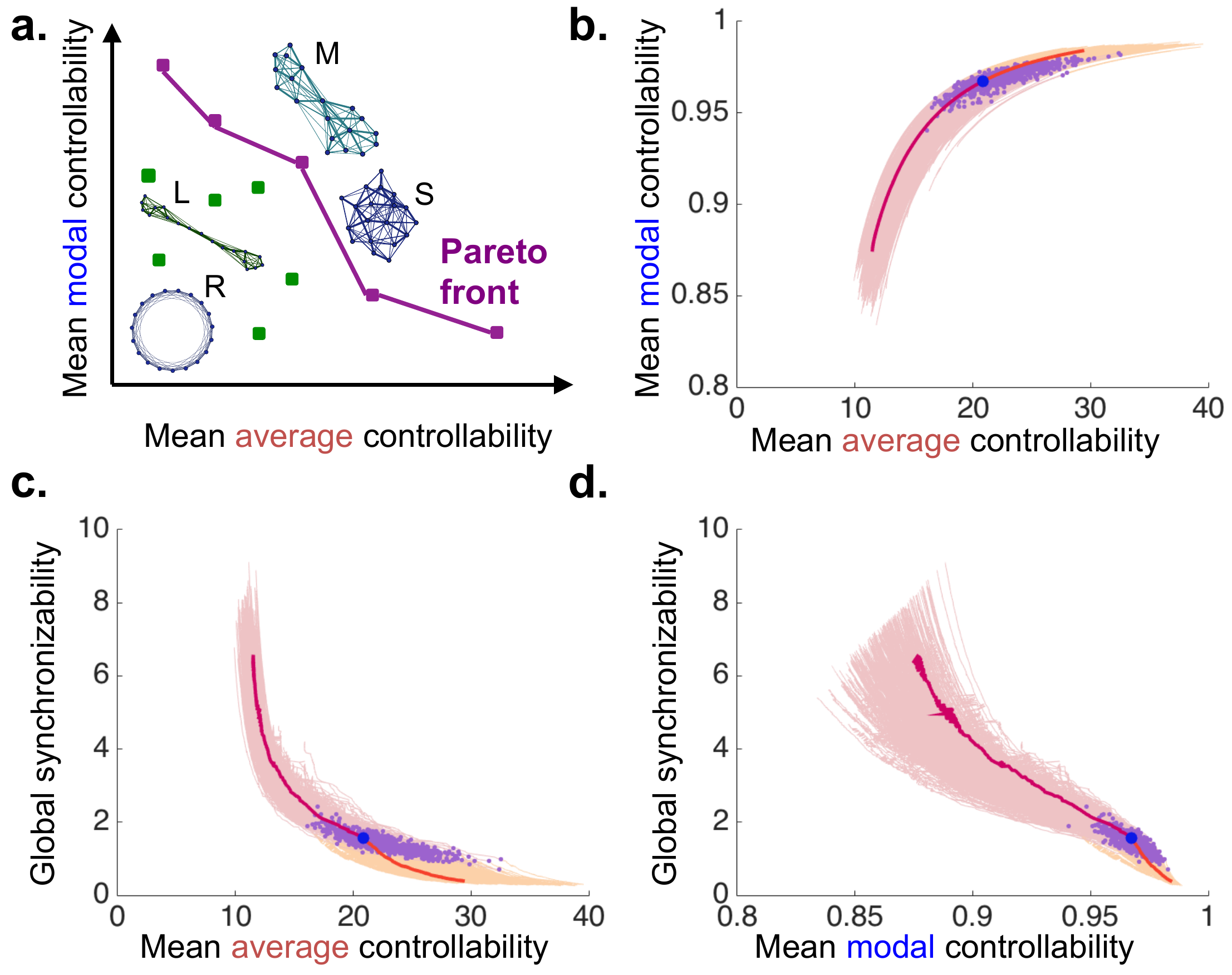}
\caption{\textbf{Brain networks are optimized for diverse dynamics over a fixed developmental time period between the age of 8 years and the age of 22 years.} \emph{(a)} Pareto optimization explores a family of networks with different topologies and hence varying mean controllability and synchronizability (a few toy models illustrate this including a ring lattice $R$, regular lattice $L$, modular network $M$, and small-world network $S$). Pareto optimal networks (purple dots) are the networks where these properties are most efficiently distributed, i.e., it is impossible to increase one property without decreasing another property -- unlike in the non-optimal networks (green dots). The boundary connecting the Pareto-optimal networks forms the Pareto front (purple line). \emph{(b, c, d)} Beginning from empirically measured brain networks of youth between the ages of 8 years and 22 years (purple dots), the authors swapped edges to modify the topology and test if the modified network advances the Pareto front. This procedure charts a course of network evolution characterized by increasingly optimal features: here the authors increased mean average controllability and mean modal controllability, and decrease global synchronizability, in 1500 edge swaps (yellow curves). For comparison, the authors also evolved the network in the opposite direction (to decrease controllability and increase synchronzability, pink curves). The trajectory for one subject (blue dot) is highlighted (orange and red). The simulated trajectories defined by the rewiring rules of increasing network controllability and decreasing synchronizability provided a suprisingly good fit to the observed data, suggesting that they are sufficient mechanisms for the empirical observations. Reprinted with permission from \cite{tang2017developmental}.} 
\label{fig:tang}
\end{figure*}

In a more recent study, Tang and colleagues study individual variation in youth by examining the white matter networks of 882 individuals between the ages of 8 yr and 22 yr \cite{tang2017developmental}. Here, the authors posited that over this developmental time period, structural brain networks become optimized for a greater diversity of neural dynamics, as instantiated by recently defined metrics of network controllability \cite{gu2015controllability}. They tested the hypothesis that an observed trajectory of network change over youth could be recapitulated by a generative model that increased average controllability (predicted ease of transitioning between nearby network states -- the level of activity in each region, across the entire brain), increased modal controllability (predicted ease of transitioning between distant network states), and decreased synchronizability (predicted capacity for global synchronization). The model was initiated with a given brain network, and then evolved \emph{in silico} according to a rewiring rule such that an existing edge was randomly chosen to take the place of an edge that did not exist, and this edge swap was retained only if the new network advanced the Pareto front, the set of all network configurations that were optimal in their tradeoff between average and modal controllability (Fig.~\ref{fig:tang}). As rewiring progressed forward in time, a course was charted in which networks increased in controllability and decreased in synchronizability; while as rewiring progressed backwards in time, networks decreased in controllability and increased in synchronizability. The simulated developmental trajectories displayed a striking similarity in functional form to the observed developmental trajectories, suggesting a possible mechanism of human brain development that preferentially optimizes dynamic network control over static network architecture.

\section*{A few relevant distinctions}

In this section, we describe a few important distinctions that are particularly relevant to the understanding and further development of generative network models for neural systems. First, we will explore the relations between generative models that seek mechanisms and explanations, and null models for statistical testing of hypotheses. Second, we will discuss the important tradeoff in sufficiency of a generative model \emph{versus} redundancy. Third, we will seek to disambiguate between inferring a possible mechanism \emph{versus} claiming proof of a mechanism. And finally, we will describe some relevant considerations when building or evaluating generative models of structural \emph{versus} functional connectivity.

\subsection*{Generative models and null models}

The stated goals of the generative modeling approach, as described in the early sections of this review, include the identification of putative mechanisms of observed network architecture, and intuitive explanations for some of the features that characterize that architecture. Yet, depending on their degree of biological realism, such models can also be used as statistical null models, potentially enabling the dismissal of a null hypothesis. In general, topological and spatially-informed null models play a critical role in network science broadly \cite{bazzi2016generative,bailey2016configuring,weidermann2015spatial}, and network neuroscience specifically \cite{samu2014influence,bassett2013robust,rubinov2011weight}. One could consider using a generative model to test the hypothesis that the topology of an empirically measured neural network was consistent with a topology of an artificial network built on a fixed set of rules or principles. In this case, one would need to be careful in the exposition of the study to distinguish between when the model was being used to propose a generative mechanism, and when the model was being used in a statistical sense to dismiss a null hypothesis.

\subsection*{Sufficiency and redundancy}

When building generative network models of empirically observed neural systems, a common observation is that the models often fit topological signatures that they were designed to fit, but rarely fit topological signatures that were not considered in the model specification \cite{klimm2014resolving}. Informally, this observation is reminiscent of the ``No free lunch'' theorem \cite{wolpert1997nofree}. However, this seeming insufficiency is not always the case \cite{rubinov2011neurobiologically,jarman2014spatially,bauer2017nonlinear}, and its inconsistent presence begs the question of what exactly makes a sufficient model. Is a sufficient network model one that can display the topological signatures it was optimized to possess (the objective function used to fit the model), or should it also predict a topological signature that was not hard coded into the objective function and/or generative algorithm? 

A complementary consideration to sufficiency is redundancy. Suppose rule \emph{a} is chosen to create topological signature \emph{1} and as a biproduct also appears to create topological signature \emph{2}. In addition, suppose that rule \emph{b} is chosen to create topological signature \emph{3} and as a byproduct also appears to create topological signature \emph{2}. Such a scenario can be quite common, as there exist whole families of graphs that display similar graph metric values \cite{costa2007characterization}, community structure \cite{onnela2012taxonomies}, controllability profiles \cite{wuyan2017benchmarking}, and homological features \cite{sizemore2017classification}. A generative model that combined rules \emph{a} and \emph{b} would appear redundant in that both rules ensured signature \emph{2}, and arguments for biological parsimony might undercut the anticipated verity of such a model. These examples illustrate that sufficiency and redundancy are important considerations in developing and evaluating generative network models of neural systems.

\subsection*{Inferring and claiming mechanism}

Suppose that one is thoroughly successful, and creates a generative model that beautifully reproduces an empirically observed network structure. Do the rules that compose the generative model provide a mechanism explaining the empirical network's architecture \cite{craver2005beyond}? Even more brazenly, can such a generative model help us to develop a theory of brain network organization and resultant behavior \cite{bassett2017tics}? In seeking answers to these questions, it is important to disambiguate between inferring a possible mechanism and claiming proof of a mechanism. If a generative network model built upon rule \emph{a} recapitulates the network structure of interest, one can say that rule \emph{a} is a possible mechanism, but one cannot claim that it is \emph{the} mechanism. To provide a more concrete example embedded in network neuroscience, let us consider the topological feature of Rentian scaling, an isometric scaling relationship between the number of processing elements and the number of connections, which is often found in systems that are built upon the principle of wiring reduction, and is observed in brain networks \cite{bassett2010efficient} as well as other transmission systems such as computer circuits \cite{landman1971pin}, transportation systems \cite{sperry2017rentian}, and vasculature \cite{papadopoulos2016embedding}. Given the scaling relationship, one might infer that the network's structure is given by a mechanism that operates uniformly across all scales such as wiring minimization. However, such an inferrence would neglect the fact that many scale-heterogeneous mechanisms also produce topological scaling relationships \cite{stumpf2012mathematics}. In future work, it will be important to concretely discuss support for possible mechanisms separately from exact claims that such mechanisms have been proven.

\subsection*{Functional connectivity and structural connectivity}

This review has focused on mostly generative models for structural networks, where links represent physical pathways among neural elements. Generative network models can also be built for functional connectivity data, with some caveats and limitations \cite{vertes2012simple,vertes2014generative,obando2017statistical}. Posited drivers of functional network organization across species include similar notions of cost-efficiency \cite{liang2017richclub,bassett2009cognitive,fornito2011genetic}, small-world architecture \cite{bettencourt2007functional}, and spatial clustering \cite{muldoon2013spatially}. However, the appropriate growth mechanisms that such generative models employ face different constraints in the functional domain than in the structural domain \cite{feldt2011dissecting}. Functional connectivity is not generated piece-by-piece, as instantiated by a discrete placement of edges in a network \cite{bazzi2015community}. Instead, functional connectivity is a consequence of dynamical processes constrained by many factors \cite{deco2009key}, including but not limited to anatomical structure \cite{honey2009predicting,goni2014resting,hermundstad2014structurally}, the activity elicited by a particular task \cite{hermundstad2013structural}, the distance between brain areas \cite{deco2009key}, genetics \cite{richiardi2015correlated, mills2017correlated, betzel2017inter}, and any stimulation or other input to the system \cite{gratton2013effect,muldoon2016stimulation}. Many good models of brain dynamics exist, ranging from the biologically realistic to the heavily idealized \cite{breakspear2017dynamic}. However, growth models built from the placement of independent edges are conceptually more appropriate for structural networks than for functional networks.

\section*{Future Directions}

\subsection*{What would a generative model accomplish?}

\begin{figure*}
	\centering
	\includegraphics[width=1.75\columnwidth]{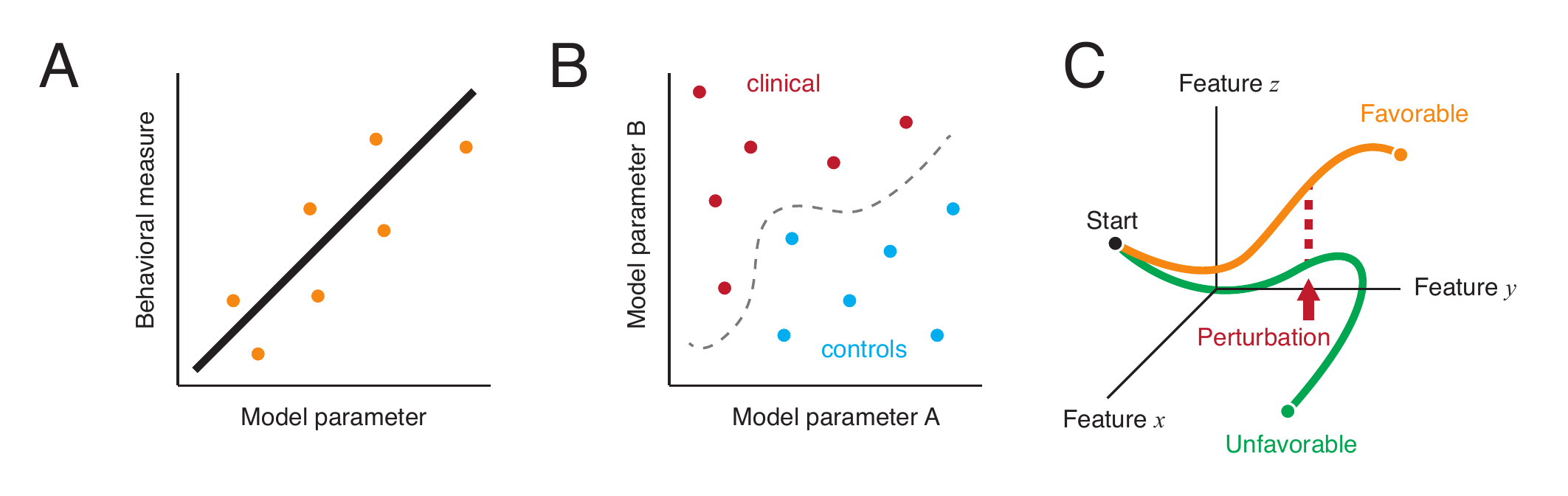}
	\caption{\textbf{Applications using generative models.} Model parameters can be fit to individual subjects and those parameters compared to some behavioral measures (\emph{A}) or used to classify different populations from one another (\emph{B}). Generative models can also be used to simulate the development of a biological neural network. These simulations can be used as forecasting devices to identify individuals at risk of developing maladaptive network topologies. They can also be used to explore possible interventions, e.g. perturbations to parameters or wiring rules, that drive an individual away from an unfavorable, maladaptive network topology towards a more favorable state.}
	\label{trajectorySchematic}
\end{figure*}

In practice, many of the current approaches for studying biological neural networks involve computing and comparing summary statistics between groups or continuously across individuals. While this approach is useful in identifying ``what'' is different, it fails to explain ``how'' those differences come to be, in the first place. In this review, we echo other recent reviews \cite{kaiser2017mechanisms,vertes2015annual} and call for a shift in emphasis away from ``fact collecting'' studies and towards uncovering the mechanisms that explain the organization of neural systems. We argue that network generative modeling represents a framework that can help us move towards addressing these lofty goals.

Suppose that -- with the right dataset and the right modeling approach -- we can devise a model that, to a reasonable approximation, can successfully mimic the growth or evolution of a real-world neural system. In other words, the model results in a network that changes over time (where time has a clear developmental or biological interpretation) and whose topology evolves in way that is consistent with known facts about the real-world growth of that network. What does having such a model buy us? On one hand we could simply maintain the \emph{status quo}, fit the model's parameters to individual subjects and compute statistical relationships between parameters and behavioral measures (Fig.~\ref{trajectorySchematic}A) using machine learning techniques to partition the model's parameter space into regions associated with clinical and control populations (Fig.~\ref{trajectorySchematic}B). While useful, these approaches are quite similar to the current state of the field.

Another more novel possibility is to use the model for disease simulation. Many psychiatric \cite{stephan2009dysconnection} and neurodegenerative diseases \cite{seeley2009neurodegenerative} are manifest at the network level in the form of miswired or dysconnected systems, but it is unclear what predisposes an individual to evolve into a disease state. The generative model can be used to propagate individuals from one time point to another and identify those that are likely to evolve into a state similar to that of the disease phenotype and perhaps likely to develop that disease. In this way, the model has a clear role as a forecaster (Fig.~\ref{trajectorySchematic}C).

Similarly, the generative model can be used to explore \emph{in silico} the effect of potential intervention strategies. We can think of biological neural networks as living in a high-dimensional space based on their topological characteristics, where some regions (of this space; not of the brain) are associated with neurological disease and considered maladaptive (and perhaps even deadly) \cite{magnin2009support, shen2010discriminative}. In this context, the generative model represents an evolution operator that propagates a network from one point to another, tracing out a trajectory through this space. If we can identify individuals who are predisposed to travel near those maladative regions, we can begin to identify perturbations -- changes to model parameters or wiring rules -- that steer those trajectories towards regions not associated with disease (Fig.~\ref{trajectorySchematic}C). These goals are in line with current theoretical work, applying tools from network control theory to neuroimaging data \cite{gu2015controllability,betzel2016optimally,iturria2017multifactorial}.

\subsection*{Dream datasets and experiments}

Generative models have clear utility in furthering our capacity to predict disease and identify the mechanisms that shape the development, growth, and evolution of biological neural networks. A major hindrance in realizing these goals, however, is the absence of data tailored for generative models. The ideal data would (i) be longitudinal, enabling one to track and incorporate individual-level changes over time in the model, and (ii) include multiple data modalities, such as functional and structural connectivity, and genetics, along with other select factors that could influence network level organization. In short, any meta-data that could theoretically be incorporated into a model would be valuable and possibly worth collecting. Ideally, these data would be acquired at the earliest possible time point \emph{in utero} \cite{heuvel2016functional} and proceed through maturity.

Clearly, collecting and curating such a dataset represents a massive undertaking. Though recent large-scale studies have made it possible to image thousands of individuals over a short period of time \cite{nooner2012nki,van2013wu,sudlow2015uk} and a small number of individuals over a long period of time \cite{laumann2015functional, braga2017parallel, gordon2017precision}, the duration and scale of a longitudinal study of the nature proposed here seems, at present, out of reach. Furthermore, the studies that have come closest to acquiring these kind of data have relied on MRI due to its non-invasive nature. However, this same advantage also limits the fidelity and kinds of data that can be acquired from an individual (e.g., region-specific gene transcription levels can only be acquired post-mortem \cite{hawrylycz2012anatomically}).

An attractive alternative, then, is to consider building generative models of data from non-human, model organisms. Not only are the lifecycles of several model organisms much shorter than that of humans (making it possible to track an individual over the course of its entire life), but new advances in network reconstruction techniques \cite{zador2012sequencing, chung2013structural, helmstaedter2013connectomic} and the ability to make recordings of activity in unprecedented detail \cite{kim2016long, eichler2017complete} ensure that any generative model will be endowed with sufficiently rich data to probe for novel wiring rules. Moreover, working with model organisms also makes it possible to collect data modalities that, otherwise, would be inaccesible, including details about gene expression \cite{lein2007genome}.

\subsection*{Increasing sophistication of generative network models}

Finally, given ideal data, there are also exciting and important future directions in increasing the mathematical sophistication of generative network models. One particularly accessible extension of current methods lies in multilayer generative network models. A multilayer network consists of multiple single-layer networks, e.g. representing a neural systems structural connectivity, functional connectivity, and gene co-expression \cite{battiston2017multilayer,bentley2016multilayer}, that are linked across layers to one another. A generative model for this type of data is one that, instead of single-layer networks, generates multilayer networks \cite{kivela2013multilayer}, and the rules of generation can apply to a single layer, to multiple layers, or to the interconnectivity between layers \cite{nicosia2017collective}. One potentially useful place to start would be to construct multilayer generative models where the neural connectivity evolves with a specific set of dynamics (or network growth rules) that are explicitly coupled to the underlying tissue growth or to the inervating vasculature growth \cite{ronellenfitsch2016global}. At the larger scale, one could also consider developing multilayer generative models that couple brain network growth with social network growth, a coupling that has recently been postulated to occur through processes of development and learning \cite{falk2017brain}. 

Indeed, it is likely that there are other ways in which our brain network topology, and changes in that topology, are coupled to our experiences. Such experiences could be defined by our environment, for example as partially stipulated by our socio-economic status \cite{ursache2016socioeconomic}, or by our practices, for example as instantiated in our practice of curiosity \cite{bassett2017curiosity}. Indeed, it is interesting to speculate that generative network models may be useful in understanding the relations between brain network architecture and the architecture of knowledge networks, which are physically instantiated in the brain \cite{constantinescu2016organizing}, as well as semantic networks \cite{huth2016natural}, which can be tuned by our attention \cite{cukur2013attention}.  Semantic networks, social networks, brain networks, vasculature networks, and tissue networks may all evolve with one another in inter-twined multilayer network systems, an understanding of any pair of which will require concerted efforts in extending the sophistication of current generative network modeling techniques.

\section*{Conclusion}

As the field of network neuroscience matures, efforts in data description and statistical characterization are being complemented by efforts to infer principles, to predict unobserved data, and to perturb the system with theoretically grounded expectations about the results of those perturbations. Generative modeling is a particularly powerful approach for moving beyond description towards prediction, mechanism, and eventually theory. In this article, we have offered a simple primer on generative models, a review of recent efforts in generative models of biological neural networks, and a discussion of current frontiers in empirical data collection and mathematical sophistication.  We look forward with anticipation to efforts in the coming years that use generative models to understand human development, and to potentially inform interventions in psychiatric disease or neurological disorders in which wiring patterns have gone awry.

\section*{Acknowledgments}

The authors thank Lia Papadopoulos and Evelyn Tang for helpful comments on earlier versions of this manuscript. This work was supported by the John D. and Catherine T. MacArthur Foundation, the Alfred P. Sloan Foundation, the Army Research Laboratory and the Army Research Office through contract numbers W911NF-10-2-0022 and W911NF-14-1-0679, the National Institute of Health (2-R01-DC-009209-11, 1R01HD086888-01, R01-MH107235, R01-MH107703, R01MH109520, 1R01NS099348 and R21-M MH-106799), the Office of Naval Research, and the National Science Foundation (BCS-1441502, CAREER PHY-1554488, BCS-1631550, and CNS-1626008).The content is solely the responsibility of the authors and does not necessarily represent the official views of any of the funding agencies.

\bibliography{biblio,biblio_dsb}

\end{document}